\documentclass[preprint,1p,11pt]{elsarticle}
\usepackage{amssymb}

\newcommand{\bfa}[1]{\mbox{\boldmath${#1}$}}
\newcommand{\stlr}[1]{\stackrel{\leftrightarrow}{#1}}
\newcommand{\stl}[1]{\stackrel{\leftarrow}{#1}}
\newcommand{\str}[1]{\stackrel{\rightarrow}{#1}}
\newcommand{\bea}{\begin{eqnarray}}
\newcommand{\eea}{\end{eqnarray}}
\newcommand{\bnn}{\begin{eqnarray*}}
\newcommand{\enn}{\end{eqnarray*}}
\newcommand{\be}{\begin{equation}}
\newcommand{\ee}{\end{equation}}

\newcommand{\nn}{\nonumber}

\journal{academic journal}
\begin{document}
\begin{frontmatter}
\title{A second order differential equation for the 
relativistic description of electrons and photons} 

\author{S. Ulrych}
\address{Institut f\"ur Theoretische Physik, Universit\"at T\"ubingen,
Auf der Morgenstelle 14, 72076 T\"ubingen, Germany}
\date{26 April 1999}

\begin{abstract}
A new relativistic description of quantum electrodynamics is presented.
Guideline of the theory is the Klein-Gordon equation, which
is reformulated to consider spin effects.
This is achieved by a
representation of relativistic vectors
with a space-time algebra made up of
Pauli matrices and hyperbolic numbers.
The algebra is used to construct the differential operator 
of the electron as well as the photon wave equation.
The properties of free electron and photon states related to 
this wave equation are investigated.
Interactions are introduced as usual with the minimal substitution of the
momentum operators. It can be shown that 
the new wave equation is equivalent to the quadratic form
of the Dirac equation. Furthermore, the Maxwell equations
can be derived from the corresponding wave equation for photons. 
\end{abstract}
\begin{keyword}
\PACS 12.20.-m \sep 03.65.Pm \sep 11.15.-q \sep 32.30.-r
\end{keyword}
\end{frontmatter}

\section{Introduction}
The Dirac equation \cite{Dir28} is considered as the 
fundamendal equation for  
the description of relativistic particles.
Theoretical calculations within quantum electrodynamics 
agree with experimental results to highest precision. 
In addition, the Dirac equation is the basis of
the theory of electroweak interactions, quantum chromodynamics and
quantum hadrodynamics. 
In all cases, the combination of the relativistic dispersion
relation with the Pauli spin matrices  
provides a natural explanation for, e.g., energy spectra, 
polarization observables, and cross sections.
Nevertheless, one has to ask whether 
there exist other possibilities  
to introduce spin in relativistic quantum physics. 

The Klein-Gordon equation seems to be
the most natural starting point for the description of 
quantum mechanical wave phenomena, but
spin is not considered in this equation.
The present work wants to show that
there exists a modification of the Klein-Gordon equation, which 
includes the relativistic spin effects. 
The equation can be obtained using an 
algebraic representation of the relativistic 
vector space, where the basis vectors are 
given by the elements of a Clifford
algebra. The importance of such Clifford algebras
in physical applications has been investigated by Hestenes \cite{Hes66}. 
In the last years this approach has become more and more
popular in the description of physical processes \cite{Ban95}.
In the present study these concepts will be applied. However,
a new relativistic algebra is introduced, which represents
a modification of the quaternion formalism.
The quaternion algebra is altered 
with the help of the hyperbolic numbers, a 
number system which has a long history \cite{Dur35,Cap41} but 
is rarely used in physical applications 
(see e.g.~\cite{Sor79,Fje86,Hes91,Kel94,Ant98}). 

The representation matrices of the modified quaternions form the
basis vectors of the relativistic vector space.
Using these vectors  
it is possible to define a differential operator which is
identical to the differential operator of the classical wave
equation but transforms formally like a spin operator.
The operator will be applied for the definition
of the electron wave equation,
which is called \textit{quantum wave equation}.

The Dirac equation is a first order differential equation,
whereas the Maxwell equations correspond to a second 
order differential equation for the photon field. 
The initial motivation of this paper was the question whether
one can find a differential operator which forms the basis 
for the description of the fundamental 
charged fermion field as well as for the 
corresponding gauge field. 
On the level of quantum electrodynamics 
such a unification is possible, since from
the quantum wave equation for photons,
constructed with the same differential operator, 
the Maxwell equations can be derived. 

In a theory for free fields the spin structure
of the quantum wave equation is not important. 
In this case the differential operator is 
identical to the differential operator of the classical wave
equation, i.e.~the mass operator of the Poincar\'e group.  
Therefore, free plane wave states of electrons and photons
are investigated within this group.
In particular, the investigation of the spin is 
needed for the construction of 
the theory. 

The introduction of interactions can be done
with the conventional method of minimal substitution, leading
to a Lagrangian which is invariant under gauge
transformations. This opens 
the possibility to compare the formalism with the 
Dirac theory. It can be shown that the quantum wave equation
corresponds to the quadratic form of the Dirac equation. 
Furthermore, the inhomogeneous terms of the
Maxwell equations are given
in the correct form. This connection supports the assumption 
that the calculation of physical processes will give similar results as 
in the conventional theories. 

The organisation of the paper is as follows.
In Sections \ref{not} and \ref{vec} hyperbolic numbers and
relativistic vectors including their Lorentz and 
Poincar\'e transformation properties
are investigated.
In Section \ref{spino1} the quantum wave equation for electrons is 
introduced. 
Plane wave states for electrons are investigated 
in Section \ref{planewave}.  
The wave equation can be used also 
for anticommuting field operators. 
This will be shown in Sections \ref{charges} and \ref{Canon}.
In Section \ref{electro}
the Maxwell equations are derived starting from 
the quantum wave equation for photons. 
The photon plane wave states are investigated in Section \ref{photop}.
Interactions between electrons
and photons are introduced and the equations
of motion of quantum electrodynamics are calculated in Section \ref{blab1}. 
In Section \ref{blab2} 
the connection with the Dirac theory is discussed.

\section{Complex hyperbolic numbers}
\label{not}
In the present investigation 
hyperbolic numbers are used for the mathematical formalation of the electron
and the photon wave equation.
Since these numbers are rarely used in
physical applications a brief introduction of
this number system is given.

In combination with the
complex numbers the hyperbolic numbers $x\in\bfa{H}$ are defined as  
\be
\label{beg}
x=x_0+jx_1\,,\quad x_0,\,x_1 \in\bfa{C}\,,
\ee
where the hyperbolic unit $j$ has the property
\be
j^2=1\,.
\ee
This leads to the following rules for the multiplication
and addition of two hyperbolic numbers
$x=x_0+jx_1$ and $y=y_0+jy_1$
\be
x+y=(x_0+y_0)+j(x_1+y_1)\,,\quad xy=(x_0y_0+x_1y_1)+j(x_0y_1+x_1y_0)\,.
\ee
Since there exist non-zero elements which have no inverse
these numbers form a commutative ring.
The hyperbolic unit $j$
provides a relation between
the hyperbolic sine, cosine
and the exponential function
\be
\cosh{\phi}+j\sinh{\phi}=e^{j\phi}\,, 
\ee
which can be derived in the same way as the
corresponding relation for
the complex numbers. 

Two conjugations will be used.
The conventional complex conjugation changes the sign of the
complex unit $i$ but leaves the hyperbolic unit $j$
unchanged 
\be
x^{*}=x_0^{*}+jx_1^{*}\,.
\ee
In addition, a hyperbolic conjugation will be introduced
which changes only the sign of the
hyperbolic unit
\be
\label{conj}
x^{-}=x_0-jx_1\,.
\ee 

The properties and definitions of the hyperbolic
numbers presented here are sufficient for the following investigations.
More informations about the hyperbolic number system
can be found e.g.~in the 
Refs.~\cite{Ban95,Dur35,Cap41,Sor79,Fje86,Hes91,Kel94,Ant98}.

\section{Relativistic vectors and the spin group}
\label{vec}
\subsection{Relativistic vectors}
\label{vectro}
The relevance of Clifford algebras for the mathematical description of
physical theories has been investigated by Hestenes \cite{Hes66}. He
reinterpretated the elements of the Pauli or the Dirac algebra
as the basis vectors of a vector space.
In the same way the present study will be 
based on such a geometric algebra.
However, a new relativistic algebra is introduced, which represents
a modification of Hamilton's quaternions.  

A contravariant Lorentz vector with the
coordinates $x^\mu=(x^0,x^i)\in \bfa{C}^4$
can be expressed as follows 
\be
\label{veco}
X=x^\mu\sigma_\mu\,.
\ee
In contrast to the formalism 
used in the context of the $SU(2)$ group, 
the basis vectors $\sigma_\mu$ are made up of the
unity and the elements of the Pauli algebra
multiplicated by the hyperbolic
unit~$j$ 
\be
\sigma_\mu=(1,j\sigma_i)\,.
\ee
Two other notations for the vector $X$ to be used in
the following are 
\be
X=x^0+j\bfa{x}=x^01+jx^i\sigma_i\,.
\ee
This means the elements of the Pauli algebra will be included in the following 
in the three-dimensional vectors $\bfa{x}=x^i\sigma_i$.
This notation has become popular in the
physical applications of Clifford algebras. The Pauli algebra
is characterized by its multiplication rules, which can be
written as 
\be
\sigma_i\sigma_j=\delta_{ij}1+i\epsilon_{ijk}\sigma^k\,.
\ee
Using the Pauli matrices as the explicit representation of $\sigma_i$, 
the vector $X$ can be expressed in terms of a
$2\times 2$ matrix according to 
\be
X=\left(\begin{array}{cc}
x^0+jx^3&jx^1-ijx^2\\
jx^1+ijx^2&x^0-jx^3\\
\end{array}\right)\,.
\ee
The formalism is not restricted to four-dimensional 
vectors. Adding a vector which is multiplicated by the factor $ij$  
an eight-dimensional multivector 
can be constructed 
\be
\label{mult}
Z=X+ijY\,.
\ee
$X$ denotes the vector contribution, whereas 
$Y$ is interpretated as a pseudovector.
This interpretation follows from a comparision with the
Maxwell equations, which are
given in Section \ref{electro}.
The eight complex coordinates of $Z$ in Eq.~(\ref{mult}) are
the maximum number of coordinates that can be 
placed in a $2\times 2$ matrix. This means, comparing 
with the sixteen-dimensional complex vector space
of the $4\times 4$ Dirac matrices, 
the new relativistic formalism
halves the number of dimensions arising from
the mathematical structure of the vector space.
In the following only the real vector
coordinates of the multivector will be considered, i.e.~the
investigation is restricted to 
the four-dimensional Minkowski space.

A scalar product between two vectors 
can be defined using the trace of the matrix $\bar{X}Y$ 
\be
\label{scalar}
\langle\,X\,\vert\, Y
\,\rangle  =\frac{1}{2}\,Tr\,(\bar{X}Y)\,.
\ee 
The symbol $\bar{X}=X^{\dagger\, {-}}$
denotes transposition, complex and 
hyperbolic conjugation of the matrix. 
$\bar{X}Y$ corresponds to a matrix multiplication
of the two $2\times 2$ matrices $\bar{X}$ and $Y$.
As stated above, the Pauli matrices can be considered
as the basis vectors of the relativistic vector space 
$\sigma_\mu\equiv e_{\mu}$.
These basis vectors form a non-cartesian orthogonal basis with respect to the
scalar product defined in Eq.~(\ref{scalar}) 
\be
\langle\,e_{\mu}\,\vert\, e_{\nu}\,\rangle=g_{\mu\nu}\,,
\ee
where the metric tensor $g_{\mu\nu}$ is a diagonal $4\times 4$ matrix
with the matrix elements 
\be
g_{\mu\nu}=\left(\begin{array}{cccc}            
1&0&0&0\\
0&-1&0&0\\
0&0&-1&0\\
0&0&0&-1\\
\end{array}\right)\,.
\ee
The metric tensor can be used as usual for raising and lowering
the indices. Eq.~(\ref{scalar}) 
can now be expressed in the conventional notation. One finds 
\be
\label{conv}
\langle\,X\,\vert\, Y\,\rangle=\langle\,e_{\mu}\,\vert\,
x^{\mu} y^\nu\,\vert\, e_{\nu}\,\rangle
=x_\mu y^\mu\,,
\ee
and the infinitesimal distance corresponds to 
\be
ds^2=
\langle\,dX\,\vert\, dX\,\rangle=dx_\mu dx^\mu\,.
\ee

As an example 
the energy-momentum vector of a free 
classical pointlike particle, moving 
with the velocity 
$\bfa{v}$ relative to the observer, 
is expressed in terms of the matrix algebra.
The relativistic momentum vector for this 
particle can be written as 
\be
\label{koko}
P=\frac{E}{c}+j\bfa{p}=mc\exp{(j\bfa{\xi})}\,,
\ee
with $c$ denoting the velocity of light,
$\bfa{\xi}$ the rapidity, $E$ the energy and
$\bfa{p}$ the momentum of the particle. 
The rapidity is defined as
\be
tanh\xi=\frac{v}{c}=\frac{pc}{E}\,,
\ee
where $\xi=|\bfa{\xi}|$ and $p=|\bfa{p}|$.
Rapidity and momentum 
point into the same direction $\bfa{n}=\bfa{v}/|\bfa{v}|$ as the velocity.

In quantum mechanics
energy and momentum are substituted 
by differential operators. With $\nabla=\partial^\mu\sigma_\mu$
the momentum operator is then given by
\bea
\label{nabo}
P=i\hbar\nabla\,.
\eea
This operator forms the basis of the 
wave equation with spin, which will be introduced in Section \ref{spino1}.
In the following $c$ and $\hbar$ will be set equal to one.

\subsection{Lorentz and Poincar\'e Transformations}
\label{trafo}
In analogy to the relation between $SU(2)$ and $SO(3)$
the transformation properties of the
vectors defined in the last subsection 
give a relation between $SO(3,1)$ and
a spin group defined as an extension of
the unitary group $SU(2)$. In the following rotations 
and boosts will be investigated.
The rotation parameters $\bfa{\theta}$ 
are defined with the conventions of Ref.~\cite{Tun85}.
The rotation of a vector has the form 
\bea
\label{rota}
&&X\,\mapsto\, X^\prime=R X\, R^\dagger\,,\nn\\
&&R=\exp{(-i\bfa{\theta}/2)}\,,
\eea
where $X=x^\mu\sigma_\mu$.
In addition, a vector can be boosted
to a different system. The boost parameters $\bfa{\xi}$ are chosen
to make the considered vector describe an object moving into the positive
direction for positive values of $\bfa{\xi}$. 
In many investigations a different sign convention is used. 
For the boosts one finds the transformation rule
\bea
\label{boost}
&&X\,\mapsto\, X^\prime= B X B^\dagger\,,\nn\\
&&B=\exp{(j\bfa{\xi}/2)}\,.
\eea
The dagger in the above equations includes only a 
hermitian conjugation and \textit{not} a
hyperbolic conjugation. For the boost
transformation one finds the relation 
$B^\dagger = B$, whereas the inverse of the
boost operator corresponds to 
$B^{-1}=\bar{B}$.
The explicit matrix representations 
of the boost matrices $B$ are
\be
B_x=\left(\begin{array}{cc}
\,\cosh{\xi^1/2}\,&\,j\sinh{\xi^1/2}\,\\
\,j\sinh{\xi^1/2}\,&\,\cosh{\xi^1/2}\,\\
\end{array}\right)
\ee
for a boost in the direction of the $x$-axis and 
\be
B_y=\left(\begin{array}{cc}
\cosh{\xi^2/2}&-ij\sinh{\xi^2/2}\\
ij\sinh{\xi^2/2}&\cosh{\xi^2/2}\\
\end{array}\right),\quad
B_z=\left(\begin{array}{cc}
e^{j\xi^3/2}&0\\
0&e^{-j\xi^3/2}\\
\end{array}\right)
\ee
for boosts along the $y$- and the $z$-axis.

To proof that these transformation matrices are
a representation of the Lorentz group the corresponding Lie algebra
has to be investigated. Before doing this, 
the three-dimensional dot and cross products will be introduced, which
are given for the three-dimensional vectors $\bfa{x}$ and $\bfa{y}$
as
\be
\label{produ}
\bfa{x}\cdot\bfa{y}=Re\left\{\bfa{x}\bfa{y}\right\}=x_iy^i\,,\quad 
\bfa{x}\times\bfa{y}=Im\left\{\bfa{x}\bfa{y}\right\}=\epsilon_{ijk}x^i
y^j\sigma^k\,.
\ee
If boosts and rotations
are combined as follows 
\bea
\label{lorentz}
&&X\,\mapsto\, X^\prime= L X L^\dagger\,,\nn\\
&&L=\exp{\left(-i(\bfa{J}\cdot\bfa{\theta}+\bfa{K}\cdot\bfa{\xi})\right)}\,,
\eea
the infinitesimal generators of these 
transformations can be identified with 
\be
\label{gener}
J_i=\frac{\sigma_i}{2}
\,,\quad
K_i= ij\frac{\sigma_i}{2}\,. 
\ee
With the commutation relations of the Pauli matrices one
can derive that the generators satisfy the  
Lie algebra of the Lorentz
group $SO(3,1)$ 
\bea
\label{commu1}
\left[\,J_i\,,\,J_j\,\right]\!\!\!\!&=&\!\!\!\!i\varepsilon
_{ijk}J^k\,,\nonumber\\
\left[K_i,\,J_j\,\right]\!\!\!\!&=&\!\!\!\!i\varepsilon_{ijk}K^k\,,\\
\left[K_i,K_j\right]\!\!\!\!&=&\!\!\!\!-i\varepsilon_{ijk}J^k\nonumber\,.
\eea
Therefore, the 
matrices $R$ and $B$ given in Eqs.~(\ref{rota}) and
(\ref{boost}) can be recognized as the
transformation matrices of the covering, i.e.~the spin group of
$SO(3,1)$.

It is possible to express
the Lorentz transformations in the conventional tensor formalism.
Using the generators of
the spin $s=\frac{1}{2}$ representation given in
Eq.~(\ref{gener})
the relativistic generalization
of the spin angular momentum operator
can be defined as
\be
\label{spinref}
S_{ij}=\epsilon_{ijk}J^{k}\,, \quad  S_{0i}=-S_{i0}=K_i\,
\ee
and the Lorentz transformations given in Eq.~(\ref{lorentz})
can be formulated according to
\be
\label{relspin}
L=\exp{(-\frac{i}{2}S_{\mu\nu}\omega^{\mu\nu})}\,.
\ee
The boost parameters $\omega^{\mu\nu}$ are
given as $\omega^{ij}=\epsilon^{ijk}\theta_k$ and
$\omega^{i0}=\xi^i$.

The relativistic orbital angular momentum can be
introduced in terms of vector operators for position and momentum
obeying the following commutation relations 
\be
[X^\mu,X^\nu]=0\,,\quad
[P^\mu,P^\nu]=0\,,\quad
[P^\mu,X^\nu]=ig^{\mu\nu}\,.
\ee
In the following the convention is used that big letters denote
operators and small letters numbers.
The relativistic orbital angular momentum is defined as
$L^{\mu\nu}=X^\mu P^\nu-X^\nu P^\mu$. 
If the operators $J_i$ and $K_i$ are given 
according to
\be
\label{ospinref}
J_i=\frac{1}{2}\epsilon_{ijk}L^{jk}\,, \quad  K_i=L_{0i}=-L_{i0}\,,
\ee
these generators satisfy the Lie algebra of the Poincar\'e group,  
i.e.~beside the relations given
in Eq.~(\ref{commu1}) one finds the following commutation relations
\bea
\label{commu2}
\left[\,J_i\,,P_0\right]\!\!\!\!&=&\!\!\!\!0\,,\nonumber\\
\left[\,J_i\,,P_j\right] \!\!\!\!&=&\!\!\!\!i\varepsilon_{ijk}P^k\,,\nonumber\\
\left[K_i,P_0\right]\!\!\!\!&=&\!\!\!\!-iP_i\,,\\
\left[K_i,P_j\right]\!\!\!\!&=&\!\!\!\!-i\delta_{ij}P_0\nonumber\,.
\eea

The Lorentz transformations corresponding to Eq.~(\ref{relspin})
can now be expressed as 
\be
L=\exp{(-\frac{i}{2}L_{\mu\nu}\omega^{\mu\nu})}\,,
\ee
The boost parameters $\omega^{\mu\nu}$ are
defined as $\omega^{ij}=\epsilon^{ijk}\theta_k$ and
$\omega^{i0}=\xi^i$. This transformation is 
acting on relativistic Hilbert space functions, which will not
be specified here further. 
To complete the transformation properties the translations
are introduced by
\be
\label{trans}
T=\exp{(-iP_\mu a^\mu)}\,.
\ee

Some remarks on the conventions should be made here. In the $SO(3)$ 
subspace of the Lorentz group, which is indicated by the roman indices,
there is no difference between 
upper and lower components, i.e.~$x^i=x_i$. The relativistic 
contravariant vector is then given as 
$x^\mu=(x^0,x^i)$ and the covariant vector as $x_\mu=(x_0,-x_i)$.
A similar convention is made for tensors. One
finds e.g.~for the zero-components of the orbital angular momentum
$L^{0\mu}=(0,L^{0i})$ and $L_{0\mu}=(0,-L_{0i})$,
where $L^{0i}=L_{0i}$.
For the relativistic spin matrices $\sigma_\mu$ a special
notation is used in which $\sigma_i$ furthermore corresponds
to a Pauli matrix.

\section{The quantum wave equation for electrons}
\label{spino1}
The Dirac equation is accepted as the fundamental relativistic equation 
for the description of fermionic particles. 
In the present work a relativistic wave equation
will be introduced, which is closely related to
the classical wave equation. 
The new equation will be called quantum wave equation.
The differential operator of the quantum wave equation
is formed by the momentum operator $P$ multiplied by
the dual operator $\bar{P}$, which is dual in the
sense that a scalar is obtained if the operator
product is inserted between two dual spinor functions.
Therefore, the following ansatz is made for the quantum wave equation 
\be
\label{equat}
P\bar{P}\psi(x)=m^2\psi(x)\,,
\ee
where $P=i\nabla$.
The differential operator $P\bar{P}$
can be replaced by $PP^{-}$ since the momentum
operator is hermitian. 
In the following investigations
there are no differences between these two choices
even if interactions are introduced. The 
wave function $\psi(x)$ has the general structure 
\be
\label{Ansatz}
\psi(x)=\varphi(x)+j\chi(x)\,,
\ee
where $\varphi(x)$ and $\chi(x)$ are two-component
spinor functions. They depend on the four space-time
coordinates $x^\mu$.
The transformation properties of the operator $\bar{P}$
can be deduced by a hermitian and hyperbolic conjugation of
the corresponding equations given in the last section.

In order to clarify the structure of the quantum wave equation,
some explicit details are presented.
The Pauli matrices in Eq.~(\ref{equat})
become apparant if one inserts
Def.~(\ref{veco})
\be
\label{seperate}
\sigma_\mu\bar{\sigma}_\nu P^\mu P^\nu\psi(x)=m^2\psi(x)\,.
\ee
The tensor $\sigma_\mu\bar{\sigma}_\nu$ 
represents the spin structure which is acting
on the spinor function. 
The explicit form is obtained by a 
matrix multiplication of the $2 \times 2$ basis matrices.
The tensor can be separated into a symmetric and an
antisymmetric contribution  
\be
\label{spindef}
\sigma_\mu\bar{\sigma}_\nu=g_{\mu\nu}-i\sigma_{\mu\nu}\,,
\ee
where $g_{\mu\nu}$ corresponds to the metric tensor and
the antisymmetric part is given by
\be
\label{sigspin}
\sigma_{\mu\nu}=
\left(\begin{array}{cccc}
0&-ij\sigma_1&-ij\sigma_2&-ij\sigma_3\\
ij\sigma_1&0&\sigma_3&-\sigma_2\\
ij\sigma_2&-\sigma_3&0&\sigma_1\\
ij\sigma_3&\sigma_2&-\sigma_1&0\\
\end{array}\right)\,.
\ee
The antisymmetric contribution $\sigma_{\mu\nu}$ 
is directly related to the relativistic
generalization of the spin angular momentum operator.
Using the generators of
the spin $s=\frac{1}{2}$ representation given in
Eq.~(\ref{spinref}) one finds
\be
\label{spinref2}
S_{\mu\nu}=\frac{\sigma_{\mu\nu}}{2}\,.
\ee
Since $P^\mu P^\nu$ is symmetric and $\sigma_{\mu\nu}=-\sigma_{\nu\mu}$
the operator $P\bar{P}$
is equivalent to
$P\bar{P}=P_\mu P^\mu$.

At this point the particular form of the differential operator
seems to be without any effect.
However, the spin information which is included in the
differential operator of the quantum wave equation becomes essential
if the momentum operators are replaced by covariant derivates.
The influence of this spin structure 
can be illustrated by the following example.
Coordinate and momentum vector satisfy the
relation
\be
X\bar{P}=X_\mu P^\mu -iS_{\mu\nu}L^{\mu\nu}\,,
\ee
where $L^{\mu\nu}=X^\mu P^\nu-X^\nu P^\mu$ corresponds to
the relativistic orbital angular momentum.

\section{Electron plane wave states}
\label{planewave}
In this section the solutions of the quantum
wave equation for electrons will be studied in the free non-interacting case. 
Due to the simplification
of the differential operator $P\bar{P}=P_\mu P^\mu$, which is 
identical with the mass operator of the Poincar\'e 
group, the solutions will be expressed in terms of
the corresponding plane wave representations. 
The section is separated into three parts.  
The first part describes how the plane
wave states can be generated. 
The second part gives additional information on
the relativistic bracket notation, which will be
applied in the following. The third 
section investigates the connection between spin and 
Pauli-Lubanski vector. 

\subsection{Induced representation method}
The irreducible representations of the
Poincar\'e group were investigated by Wigner \cite{Wig39}.
He found that the plane wave states 
are labelled by the mass $m$ and the spin $s$. 
In the present work these states will be generated
with the induced representation
method, which is described e.g.~in
Ref.~\cite{Tun85}.
In this method a state vector 
is defined within the little group of the Poincar\'e group,
i.e.~the subgroup that leaves a particular 
standard vector invariant. 
An arbitrary 
state is then generated with the remaining 
transformations which were not considered in the little group. 
In the following 
the transformation rules of Section \ref{trafo}
will be applied. 

For electrons $(m^2>0)$ one can choose the
standard vector $p^\mu_t=(m,0,0,0)$. 
The little group of this standard vector is $SO(3)$. 
The explicit representation
of the spin $s=\frac{1}{2}$ states 
is given by the Pauli spinor, which will be denoted by
$\vert\,\lambda\,\rangle=\chi_{\lambda}$.
The polarization is chosen along the z-axis. 
For the description of the mass quantum number
$m$ a ket $\vert\, P_t\,\rangle$ is introduced. 
By doing this, the translations, as one group of the remaining
transformations, are taken into account. 
The properties of the momentum
kets will be investigated separately in the
next section.
Note, that though the kets are represented by
$\vert\, P_t\,\rangle=\vert\, p_t^\mu\sigma_\mu\,\rangle$, they
correspond to usual Hilbert space elements which depend
only on the four-momentum $p_t^\mu$. 
One therefore starts with the following state, which corresponds to
a $(m,s)$-representation of the Poincar\'e group 
\be
\label{starts1}
\vert\,\lambda\,\rangle\otimes\vert\, P_t\,\rangle
=\vert\,\lambda\, P_t\,\rangle\,.
\ee
Now, the boosts, as the last group of the remaining transformations, 
are acting on this state 
according to
\be
\label{starts2}
D(B)\vert\,\lambda\, P_t\,\rangle=
B\chi_{\lambda}\,\vert\,B P_tB^\dagger\,\rangle\,,
\ee
where $B$ has been defined in Eq.~(\ref{boost}).
Since the boost transforms from the rest frame to a particular frame, in
which the state is described by the 
momentum $P=B P_tB^\dagger$, the boost parameters can be
identified with the rapidity. 
With this information it is possible to calculate
the explicit form of the relativistic spinor.
In analogy to the Dirac formalism one can introduce
the notation \cite{Bjo65}
\be
\label{begin}
u(\bfa{p},\lambda)
=B\chi_{\lambda}\,.
\ee 
Explicitly the boost matrix can be written as 
\be
B=\exp{(j\bfa{\xi}/2)}=\cosh{\xi/2}+j\bfa{n}\sinh{
\xi/2}\,,
\ee
where the rapidity $\xi$ satisfies the following
relations 
\be
\cosh{\xi/2}=\sqrt{\frac{p^0+m}{2m}}\,,\quad 
\sinh{\xi/2}=\sqrt{\frac{p^0-m}{2m}}\,.
\ee
Inserting these results into Eq.~(\ref{begin})
the spinor is given by 
\be
\label{spnra}
u(\bfa{p},\lambda)=\sqrt{\frac{p^0+m}{2m}}\left(1
+\frac{j\bfa{p}}
{p^0+m}\right)\chi_{\lambda}\,.
\ee

The antiparticle spinor is constructed in  
analogy to the Dirac theory, where upper and
lower components are interchanged compared to the particle spinor. 
In the formalism presented here, 
this can be done by multiplying the particle spinor by the 
hyperbolic unit $j$, i.e.~$v(\bfa{p},\lambda)=ju(\bfa{p},\lambda)$.
Therefore, one can write 
\be
\label{antis}
v(\bfa{p},\lambda)=\sqrt{\frac{p^0+m}{2m}}\left( \frac{\bfa{p}}
{p^0+m} +j\right)\chi_{\lambda}\,.
\ee
The normalization and the completeness of the plane wave states 
can be calculated  
as follows
\bea
\label{norma}
\langle u(\bfa{p},\lambda)\vert u(\bfa{p},\lambda^\prime)\rangle\!\!\!\!&=&\!\!\!\!
+\delta_{\lambda\lambda^\prime}\,,\nonumber\\
\langle v(\bfa{p},\lambda)\vert v(\bfa{p},\lambda^\prime)\rangle\!\!\!\!&=&\!\!\!\!-\delta_{\lambda\lambda^\prime}\,.
\eea
To explain this notation the
scalar product of two plane wave spinors is
determined explicitly  
\be
\label{real}
\langle u(\bfa{p},\lambda)\vert u(\bfa{p},\lambda^\prime)\rangle=
\bar{u}(\bfa{p},\lambda)u(\bfa{p},\lambda^\prime)=
\chi_{\lambda}^{\dagger} \bar{B}B\chi_{\lambda^\prime}=
\delta_{\lambda\lambda^\prime}\,.
\ee
The antiparticle spinor is normalized to $-1$, which is due
do the additional $j$-factor and its hyperbolic conjugated counterpart
$j^-=-j$.
Note, that there is no explicit 
orthogonality between particle and antiparticle spinor.
However, if the spinors appear in combination 
with the momentum kets of the states, 
orthogonality of particle and antiparticle states
will be restored. This can be checked with the
relations given in the following sections.

For the completeness one finds
\be
\label{compla}
\sum_{\lambda}\,
\vert u(\bfa{p},\lambda)\rangle\langle u(\bfa{p},\lambda)\vert=
-\sum_{\lambda}\vert v(\bfa{p},\lambda)\rangle
\langle v(\bfa{p},\lambda)\vert\,=1\,.
\ee
Together with bras and kets for the momentum these expressions correspond 
to negative and positive energy projectors, respectively.
In contrast to the appropriate expressions in the Dirac
theory, they have a simple matrix structure and no momentum
dependence.

The transformation properties of the states
are  
\bea
\label{trafog}
D(L)\,\vert\,u(\bfa{p},\lambda)\, P\,\rangle\!\!\!\!&=&\!\!\!\!
(Lu(\bfa{p},\lambda))\,\vert L PL^\dagger\,\rangle\,,
\\ 
D(T)\,\vert\,u(\bfa{p},\lambda)\, P\,\rangle\!\!\!\!&=&\!\!\!\!
u(\bfa{p},\lambda)\,\exp{(-ip_\mu a^\mu)}\vert\, P\,\rangle\,,
\nonumber
\eea
where the Lorentz transformations are performed with the
matrix $L$ of Eq.~(\ref{lorentz}).
The notation presented here was introduced in analogy to 
the Dirac formalism and it therefore adopted the expression
$u(\bfa{p},\lambda)$. In the following the
abstract notation $\vert\,q\, \lambda\,P\,\rangle$
will be used, where a quantum number denoting a particle with 
$q=1$ and an antiparticle with $q=-1$ is introduced. 
The explicit form of these states is
\be
\vert+\lambda\, P\,\rangle=\vert\, u(\bfa{p},\lambda)\,P\,\rangle\,,
\quad \vert-\lambda\, P\,\rangle=
\vert \,v(\bfa{p},\lambda)\,-\!\!P\,\rangle \,.
\ee
Again the antiparticle state is introduced
by definition. A proper derivation of these states requires 
the investigation of the discrete Lorentz transformations.

\subsection{The relativistic $\vert\, P \,\rangle$ and $\vert\, X \,\rangle$ 
representations}
\label{represent}
In the last subsection the kets
$\vert\, P \,\rangle$ and $\vert\, X \,\rangle$
for relativistic Hilbert space elements
related to a $(m^2>0)$ representation of the Poincar\'e
group were introduced. In the following the 
normalization and completeness of the states will be summarized, where
the description follows the covariant conventions 
given e.g.~in Refs.~\cite{Tun85,Ryd85}. 
The spin degrees of freedom will be considered separately in
the next section.

The normalization and completeness of the states
is expressed in an abstract notation. The
normalization is given by 
\be
\label{normket}
\langle \, P \,\vert\, P^\prime \,\rangle=\delta(P-P^\prime)\,,
\quad \langle \, X \,\vert\, X^\prime \,\rangle=\delta(X-X^\prime)\,,
\ee
and the completeness can be written as
\be
\label{compket}
\int\! dP\,\vert \, P \,\rangle\langle\, P \,\vert=1\,,
\quad \int\! dX\,\vert \, X \,\rangle\langle\, X \,\vert=1\,.
\ee
The dependence on the mass
$m$ is suppressed in the bras and kets.
The explicit covariant representation of the momentum delta function is given 
\cite{Tun85,Ryd85} according to 
\be
\delta(P-P^\prime)=(2\pi)^3\,2p^0\,\delta^3(\bfa{p}-\bfa{p}^{\prime})\,,
\ee
where $p^0=\sqrt{\bfa{p}^2+m^2}$ is fixed by the mass shell
condition.
The integration over $dP$ corresponds to
\be
\int\! dP\,=\int\! \frac{d^3\bfa{p}}{(2\pi)^3\,2p^0}\,,
\ee
with $d^3\bfa{p}=dp^1dp^2dp^3$.

Using these representations the abstract completeness and 
normalization of Eqs.~(\ref{normket}) and (\ref{compket})
are consistent in momentum space. In coordinate space the situation is
more complicated.
For the unity expressed in terms of bras and kets 
one can define
\be
\int\! dX\,\vert \, X \,\rangle\langle\, X \,\vert=
\int\! d^3\bfa{x}\,\vert \, X \,\rangle\!
\stlr{P^0}\!\langle\, X \,\vert\,,
\ee
where the abbreviation
\be
A\stackrel{\leftrightarrow}{P^\mu}B=A(P^\mu B)-(P^\mu A)B
\ee
has been introduced.
To investigate the properties of the
delta function in coordinate space the 
representation of the momentum ket is needed 
\be
\langle \, X \,\vert\, P \,\rangle=\exp{(-i p_\mu x^\mu)}\,.
\ee
With this expression the covariant 
delta function can be calculated using
the completeness of the momentum kets 
\be
\label{deltax}
\delta(X-X^\prime)=\int\! dP\,
\langle \, X \,\vert\, P \,\rangle\langle\, P \,\vert\, X^\prime \,\rangle
=\int\! \frac{d^3\bfa{p}}{(2\pi)^3\,2p^0}\exp{(-i p_\mu(x^\mu-
x^{\mu\prime}))}\,.
\ee
The delta function is therefore equivalent to the 
positive frequency part $\Delta_+(x-x^\prime)$ of the
invariant function $i\Delta(x-x^\prime)$ given in Ref.~\cite{Bjo65}. 
This reflects the fact that these considerations are
restricted to positive energy states.
The action of the delta function applied to
the state $\langle \, X \,\vert\, P \,\rangle$ 
is as follows  
\bea
\label{delta}
\int\! dX \delta(X^\prime-X) \langle \, X \,\vert\, P \,\rangle
\!\!\!\!&=&\!\!\!\!\int\! dP^\prime\int\! dX\,
\langle \, X^\prime \,\vert\, P^\prime \,\rangle
\langle\, P^\prime \,\vert\, X \,\rangle 
\langle \, X \,\vert\, P \,\rangle\nonumber\\
\!\!\!\!&=&\!\!\!\!\int\! \frac{d^3\bfa{p}^\prime}{(2\pi)^3\,2p^{0\prime}}
\int\! d^3\bfa{x}\,e^{-i p_\mu
^\prime(x^{\mu\prime}-x^\mu)}\stlr{P^0}e^{-i p_{\mu} x^\mu}=\langle \, X^\prime \,\vert\, P \,\rangle\,.
\eea
This shows that the explicit representations are 
consistent with the abstract completeness 
given in Eq.~(\ref{compket}). 

Using the relations given above one can
transform a momentum
dependent function $\psi(p)=\langle \, P \,\vert\, \psi \,\rangle$,
which is given on the mass shell, into 
the coordinate space according to
\be
\psi(x)=\int\! \frac{d^3\bfa{p}}{(2\pi)^3\,2p^0}
e^{-i p_{\mu} x^\mu}\psi(p)\,,
\ee
whereas in the opposite direction the 
transformation has the form
\be
\psi(p)=\int\! d^3\bfa{x}e^{i p_{\mu} x^\mu}
(p^0+i\partial^0)\psi(x)\,.
\ee
In the next sections the notation presented here will be applied e.g.~in
the second quantization of the electron field
on the equal-time plane.

\subsection{The Pauli-Lubanski vector and spin operators}
\label{subspin}
To show that the plane wave states derived 
in the first part of this section
correspond to an irreducible representation
of the Poincar\'e group, the connection of these states with the second
Casimir operator, the Pauli-Lubanski
vector, will be investigated. 
Since the spin $s=\frac{1}{2}$ representation is considered, the
spin angular momentum operators 
of Eq.~(\ref{spinref}) will be used for the definition of the
Pauli-Lubanski vector 
\be
W^\mu=\frac{1}{2}\epsilon^{\mu\rho\sigma\nu}S_{\rho\sigma}P_\nu=
\tilde{S}^{\mu\nu}P_\nu\,,
\ee
where $\tilde{S}^{\mu\nu}$ is the dual tensor of the relativistic
angular momentum tensor. It is interesting that the
dual tensor can also be calculated using the
relation $\tilde{S}^{\mu\nu}=-ijS^{\mu\nu}$, which can be derived
from Eqs.~(\ref{sigspin}) and (\ref{spinref2}).  
In the notation of Eq.~(\ref{veco}) 
the Pauli-Lubanski vector has the form 
\be
\label{Luba3}
W=-\bfa{J}\cdot\bfa{P}-j(\bfa{J}P^0+\bfa{K}\times\bfa{P})\,,
\ee
where in the expression $\bfa{J}=J^i\sigma_i$ 
the Pauli matrices in $J^i=\sigma^i/2$ are
coordinates of the Pauli matrices $\sigma_i$ in the basis.

With the Pauli-Lubanski vector                    
the relativistic spin operators can be defined,
where the investigation follows 
the methods given in \cite{Mic59,Wig60}.
One chooses a set of four orthogonal vectors $n^{(\nu)}$
satisfying the relation
\be
n_{\rho}^{(\mu)} n^{\rho(\nu)}=g^{\mu\nu}\,.
\ee
Using these vectors the spin operators are defined  
according to
\be
S^{i}=\frac{1}{m}W_\mu n^{\mu(i)}\,.
\ee
If the plane wave states derived in the first part of this section
shall be eigenstates of the spin operators, 
a particular set of 
orthogonal vectors $n^{\mu(\nu)}=(n^{0(\nu)},n^{k(\nu)})$ 
has to be introduced
\be
n^{\mu(0)}=\left(\frac{P^0}{m},\frac{P^k}{m}\right),\quad 
n^{\mu(i)}=\left(\frac{P^i}{m},\delta^{ki}+\frac{P^kP^i}{m(P^0+m)}
\right).
\ee
In the rest frame, these vectors reduce to the canonical orthogonal system.
The three spin operators can be regarded as the relativistic
generalization of the non-relativistic spin operators. 
They can be expressed as ($\bfa{S}=S^i\sigma_i$)  
\be
\bfa{S}=\frac{1}{m}\left(\bfa{J}P^0+\bfa{K}\times\bfa{P}-(\bfa{J}\cdot\bfa{P})
\frac{\bfa{P}}{P^0+m}\right).
\ee
For the relativistic 
spin operators one finds $\bfa{S}^2=s(s+1)=-W_\mu W^\mu/m^2$. 
The operators satisfy the 
commutation relations of the little group $SO(3)$. 
The third component of the spin vector can be used to
characterize the polarization. 

The spin operators were constructed in that way, that they
coincide with boosted generators $J^i$, where the boost matrices
are acting on the coordinates of $\bfa{J}=J^i\sigma_i$ 
\be
\label{shref}
S^i=BJ^i\bar{B}\,.
\ee
As well as in the derivation of the plane wave states,
the boost parameters in $B$ have to be 
identified with the rapidity of the state vector.
From the above equation follows that under
arbitrary Lorentz transformations 
the spin operators have to transform according to 
\be
\label{spintrans}
S^i\,\mapsto\, S^{i\prime}=LS^i\bar{L}\,,
\ee
where $L$ corresponds to the Lorentz tranformation matrix given
in Eq.~(\ref{lorentz}).
From Eq.~(\ref{shref}) one can 
deduce that it is sufficient to define the spin in the rest frame
of the state, according to the non-relativistic description 
with the non-relativistic spin operators $J^i=\sigma^i/2$, whereas 
the boost operators $K^i=ijJ^i$
are not used to characterize the single-particle state. 
Vector products of the form $A\bar{B}$,
with two arbitrary vectors $A$ and $B$,
transform in the same way as the spin operators
\be
(A\bar{B})\,\mapsto \,(A\bar{B})
^\prime=L(A\bar{B})\bar{L}\,.
\ee
Therefore, $P\bar{P}$ was chosen as the
differential operator of the quantum wave equation.
This guarantees that the operator, which is acting between
two spinor functions, shows the correct
transformation property.

Now, the properties of the positive energy states
can be summarized. 
The plane wave states for positive energies are eigenstates of the four 
operators $\{P_\mu P^\mu, P^\mu, W_\mu W^\mu, S^3\}$
and satisfy the relations
\bea
P_\mu P^\mu\,\vert+\lambda\, P\,\rangle
\!\!\!\!&=&\!\!\!\!m^2\,\vert+\lambda\, P\,\rangle\,,
\nonumber\\
P^\mu \,\vert+\lambda\, P\,\rangle
\!\!\!\!&=&\!\!\!\!p^\mu \,\vert+\lambda\, P\,\rangle\,,\nonumber\\
W_\mu W^\mu\,\vert+\lambda\, P\,\rangle
\!\!\!\!&=&\!\!\!\!-m^2s(s+1)\,\vert+\lambda\, P\,\rangle\,,
\\
S^3\,\vert+\lambda\, P\,\rangle
\!\!\!\!&=&\!\!\!\!\lambda\,\vert+\lambda\, P\,\rangle
\nonumber\,.
\eea
Together with the negative energy states 
one finds the following
normalization
\be
\langle\,q\,\lambda\, P\,\vert\,q^\prime\lambda^\prime P
^\prime\rangle=\delta_{qq^\prime}\,\delta_{\lambda\lambda^\prime}
\,\delta(P-P^\prime)
\ee
and completeness
\be
\sum_\lambda\int \! dP\,\vert+\lambda\, P\,\rangle
\langle\,+\,\lambda\, P\,\vert=1\,,\quad
\sum_\lambda\int \! dP\,\vert\,-\,\lambda\, P\,\rangle
\langle\,-\,\lambda\, P\,\vert=1\,,
\ee
where the completeness is restricted to the
subspaces of positive and negative energy, respectively. 
With these states the solution of
the quantum wave equation $\psi(x)$ can be expressed
as the following plane wave expansion
\begin{eqnarray}
\label{solute}
\psi(x)\!\!\!\!&=&\!\!\!\!
\sum_\lambda\int \! dP\,\left(\langle\,X^{}
\,\vert+\lambda\, P\,\rangle\langle\,+\,\lambda\, P\,\vert\,\psi\,\rangle
+\langle\,X\,\vert-\lambda\, P\,\rangle
\langle\,-\,\lambda\, P\,\vert\,\psi\,\rangle\right)\nonumber\\
\!\!\!\!&=&\!\!\!\!\sum_\lambda\int \! \frac{d^3\bfa{p}}{(2\pi)^32p^0}\,\left(
u(\bfa{p},\lambda)e^{-ip_\mu x^\mu}\,b(p,\lambda)+
v(\bfa{p},\lambda)e^{ip_\mu x^\mu}\,\bar{d}(p,\lambda)\right)\,.
\end{eqnarray}
At this point the solution is 
understood as an expansion of a single particle wave function.

\section{Variational principle and current conservation}
\label{charges}
A second order differential equation is expected to be
unsuitable for the description of fermionic quantum fields. 
Therefore, some quantum field theoretical
aspects will be investigated in the next two sections.
In this section a Lagrangian for a free non-interacting
theory will be introduced which leads to the quantum wave equation
using standard variational techniques.
The Lagrange equations, currents, and conserved
quantities will be derived.  

The initial point of this investigation is a Lagrangian of the form
\be
\label{Lagr} 
{\cal{L}}(x)=
\bar{\psi}(x)P\bar{P}
\psi(x)
-m^2\bar{\psi}(x)
\psi(x)\,,
\ee
where both momentum vectors are acting to the right side. 
The Lagrangian has the property of being invariant under
hermitian including
hyperbolic conjugation, i.e.~${\cal{L}}(x)=\bar{{\cal{L}}}(x)$.
Note, that the wave functions are two-component
spinor functions and a product of two 
spinorfunction $\bar{\psi}\psi\equiv\bar{\psi}_i
\psi_i$ implies 
a contraction of the two-component field functions.
The same convention is used in expressions like the
following 
\be
\frac{\partial{\cal{L}}}{\partial(\partial_{\mu}\psi)}
\delta(\partial_{\mu}\psi)
\equiv
\frac{\partial{\cal{L}}}{\partial(\partial_{\mu}\psi_i)}
\delta(\partial_{\mu}\psi_i)\,.
\ee 
The equations of motion, currents, and conserved
quantities can be derived as usual from a variational principle which
is applied to the action.
The investigation presented here follows the
common treatment of this topic, which can be found e.g.~in Ref.~\cite{Ryd85}.
Therefore, only the main points shall be discussed briefly.
The action is defined as
\be
S=\int\!d^4x\, {\cal{L}}(x)\,,
\ee
where 
$d^4x=dx^0dx^1dx^2dx^3$. 
${\cal{L}}(x)$ is a function of 
${\cal{L}}(\psi,\bar{\psi},\partial^\mu\psi,\partial^\mu
\bar{\psi},x^\mu)$. The variation 
of this 
Lagrangian is then given by
\be
\delta{\cal{L}}=\frac{\partial{\cal{L}}}{\partial{\psi}}
\delta\psi+\frac{\partial{\cal{L}}}{\partial(\partial_{\mu}\psi)}
\delta(\partial_{\mu}\psi)+
\delta\bar{\psi}\frac{\partial{\cal{L}}}{\partial\bar{\psi}}
+\delta(\partial_{\mu}\bar{\psi})
\frac{\partial{\cal{L}}}{\partial(\partial_\mu\bar{\psi})}
+(\partial_{\mu}{\cal{L}})\delta x^\mu\,,
\ee
where the variation of the coordinates and the
variation of the wave function is defined
as
\bea
x^\mu\!\!&\mapsto&\!\! x^{\prime\mu}=x^\mu+\delta x^\mu\,,\nonumber\\
\psi(x)\!\!&\mapsto&\!\! \psi^\prime(x)=\psi(x)+\delta\psi(x)\,.
\eea
Both variations vanish on a boundary $\partial R$.
Using $\delta{\cal{L}}$ the variation of the action $\delta S$ 
can be calculated. Applying the principle
of least action, $\delta S=0$, to this expression 
one obtains the Lagrange equations   
\be
\label{varaction}
\frac{\partial{\cal{L}}}{\partial{\psi}}-\partial_{\mu}
\frac{\partial{\cal{L}}}{\partial(\partial_{\mu}\psi)}=0\,,\quad
\frac{\partial{\cal{L}}}{\partial\bar{\psi}}
-\partial_{\mu}
\frac{\partial{\cal{L}}}{\partial(\partial_\mu\bar{\psi})}=0\,.
\ee
From the surface terms of $\delta S$ one can 
derive the conserved current, which has the form
\be
\label{curr0}
J^\mu=
\frac{\partial{\cal{L}}}{\partial(\partial_{\mu}\psi)}\Delta\psi
+\Delta\bar{\psi}
\frac{\partial{\cal{L}}}{\partial(\partial_\mu\bar{\psi})}
-\theta^{\mu\nu}\delta x_\nu\,.
\ee
$\Delta\psi$ corresponds to the total variaton $\psi^\prime(x^\prime)
=\psi(x)+\Delta\psi(x)$
and $\theta^{\mu\nu}$ denotes the energy momentum tensor 
defined as
\be
\theta^{\mu\nu}=
\frac{\partial{\cal{L}}}{\partial(\partial_{\mu}\psi)}(\partial^{\nu}\psi)
+(\partial^{\nu}\bar{\psi})
\frac{\partial{\cal{L}}}{\partial(\partial_\mu\bar{\psi})}
-g^{\mu\nu}{\cal{L}}\,.
\ee
The quantum wave equation can now be
derived with the help of the Lagrange equations (\ref{varaction}), where 
it is useful to rewrite the Lagrangian  
in analogy to Eq.~(\ref{seperate}) to  
separate the dynamical variables from the
basis matrices.

The currents for 
global phase transformations and translations can be calculated with  
the definition of the current given in Eq.~(\ref{curr0}).
A global phase transformation of the wave function
is expressed as 
\be
\psi(x)\,\mapsto\,\psi^\prime(x)=e^{-i\Lambda}\psi(x)\,.
\ee
This leads to the total field variations
\be
\Delta \psi=-i\Lambda\psi\,, \quad\Delta\bar{\psi}=i
\bar{\psi}\Lambda\,
\ee
and to $\delta x^\nu=0$.
Using the relation $P\bar{P}=
P_\mu P^\mu$ the current can be calculated as          
\be
\label{curr2}
J^\mu(x)=\bar{\psi}(x)\stlr{P^\mu}\psi(x)\,.
\ee
The following sections will show that if interactions are
considered the spin
structure becomes important and the appropriate
expression for the current will be more
complicated. 

With the formalism developed in Section
\ref{represent} one can relate 
the charge, defined on the equal-time plane, with a scalar product in the 
relativistic Hilbert space 
\be
\label{charge3}
\langle\,\psi\,\vert\,\phi\,\rangle=\int\!dX\,
\langle\,\psi\,\vert\,X\,\rangle\langle\,X\,\vert\,\phi\,\rangle=
\int\!d^3\bfa{x}\,\bar{\psi}(x)\stlr{P^0}\phi(x)\,.
\ee
This yields the relation
\be
\label{charge}
Q=\langle\,\psi\,\vert\,\psi\,\rangle=
\int\!dX\,\langle\,\psi\,\vert\,X\,\rangle\langle\,X\,\vert\,\psi\,\rangle=
\int\! d^3\bfa{x}J^0(x)\,.            
\ee

The four-momentum of a classical
field configuration is related to the 
translations, which are characterized by the following variations  
\be
\delta x^\mu= \epsilon^\mu\,, \quad \Delta \psi=0\,.  
\ee
The conserved current corresponds to  
the energy-momentum tensor, which has in the case of
free electrons the structure 
\be
\label{sta}
\theta^{\mu\nu}(x)=-(P^\mu \bar{\psi}(x))
(P^\nu{\psi(x)})-
(P^\nu \bar{\psi}(x))
(P^\mu{\psi(x)})
-g^{\mu\nu}{\cal{L}}(x)\,,
\ee
where $\theta^{\mu\nu}\!=\!\bar{\theta}^{\mu\nu}\!=\!\theta^{\nu\mu}$.
With this expression one can define 
the four-momentum of the field configuration according to 
\be
\label{energy1}
P^\mu=
\int\!d^3\bfa{x}\,\theta^{0\mu}(x)\,.
\ee

A certain amount of energy is able to create
electron-positron pairs, i.e.~to bring an electron from a negative 
energy state to a positive energy state. The negative
energy spectrum, which is naturally included
in a relativistic treatment of physics, 
has the consequence that a relativistic quantum theory
can only be constructed within a many-particle framework.
One possibility to go in this direction 
is to quantize the fields, i.e.~to consider them 
as operator valued quantum fields 
which are acting between many-particle states.
This procedure is called second quantization and 
will be performed in the next section. 

\section{Canonical quantization of free electrons}
\label{Canon}
In contrast to common belief
the canonical quantization is straightforward in a
theory of non-interacting electrons 
based on the Klein-Gordon equation.
The standard techniques of the
canonical quantization \cite{Bjo65} and the partial wave expansion
derived in the foregoing sections will be used
to demonstrate this.

The Lagrangian given in Eq.~(\ref{Lagr})
is equivalent to the Lagrangian of the Klein-Gordon equation
\be
{\cal{L}}=
\bar{\psi}(x)P_\mu P^\mu
\psi(x)-m^2\bar{\psi}(x)\psi(x)\,.
\ee
For the second quantization the conjugated momentum will be introduced
as usual 
\be
\pi(x)=\frac{\partial{\cal{L}}}{\partial(\partial_0\psi(x))}
=\partial^0\bar{\psi}(x)\,.
\ee
The field quantization follows from implying
equal-time anticommutation relations on the field variables
\bea
\label{anti}
\left.\{\stackrel{}{\psi}_i\!(x),\pi_j(y)\}\right|_{x^0=y^0}
\!\!\!\!&=&\!\!\!\!+i\delta_{ij}\delta^3(\bfa{x}-\bfa{y})\,,\\
\left.\{\bar{\psi}_i(x),\bar{\rule[6.5pt]{0pt}{1pt}\pi}
_j(y)\}\right|_{x^0=y^0}
\!\!\!\!&=&\!\!\!\!-i\delta_{ij}\delta^3(\bfa{x}-\bfa{y})\,.\nonumber
\eea
Subscripts where introduced to indicate the two-component
structure of the fields.

One can show that these relations are leading to the
Heisenberg equations of motion
\be
[\hat{P}^\mu,\psi(x)]=-i\partial^\mu\psi(x)\,,\quad
[\hat{P}^\mu,\bar{\psi}(x)]=-i\partial^\mu\bar{\psi}(x)\,,
\ee
where the momentum of the field configuration 
$P^\mu$ given in Eq.~(\ref{energy1}) has been changed  
to a many-particle operator $\hat{P}^\mu$ including the operator valued fields
$\psi(x)$ and $\bar{\psi}(x)$.

From the plane wave expansion of $\psi$ given in Eq.~(\ref{solute}) 
one can project out the amplitudes 
$b(P,\lambda)$ and $d(P,\lambda)$ 
according to
\bea
b(p,\lambda)\equiv\langle\,+\,\lambda\, P\,\vert\,\psi\,\rangle
\!\!\!\!&=&\!\!\!\!\int\!dX\,\langle\,+\,\lambda\, P\,\vert\,X\,\rangle
\langle\, X\,\vert\,\psi\,\rangle\nonumber\\
&\equiv&\!\!
\int\! d^3\bfa{x}\,e^{i p_{\mu} x^\mu}
\bar{u}(\bfa{p},\lambda)\stlr{P^0}
\psi(x)\,,\nonumber\\
d(p,\lambda)\equiv\langle\,\psi\,\vert -\lambda\, P\,\rangle
\!\!\!\!&=&\!\!\!\!\int\!dX\,\langle\,\psi\,\vert\,X\,\rangle
\langle\, X\,\vert-\lambda\, P\,\rangle\nonumber\\
&\equiv&\!\!\int\! d^3\bfa{x}\,\bar{\psi}(x)
\stlr{P^0}v(\bfa{p},\lambda)
e^{i p_{\mu} x^\mu}\,.
\eea
The two-component spinor functions
$\bar{u}(\bfa{p},\lambda)$ and $\psi(x)$ 
have to be contracted according to the convention
introduced in the last section.
Using the anticommutation relations for the fermion
fields one can derive the anticommutators
\be
\label{commu}
\{b(p,\lambda),\bar{b}(p^\prime,\lambda^\prime)\}=
\{d(p,\lambda),\bar{d}(p^\prime,\lambda^\prime)\}=
\delta_{\lambda\lambda^\prime}\,\delta(P-P^\prime)
\,.\nonumber
\ee
The many-particle operators $\hat{Q}$ and
$\hat{P}^\mu$ 
given in Eqs.~(\ref{charge}) and
(\ref{energy1}) can be expressed in terms of the operator valued 
amplitudes $b(P,\lambda)$ and $d(P,\lambda)$ according to 
\bea
\label{qcharg}
\hat{Q}\!\!\!\!&=&\!\!\!\!
\int\! d^3\bfa{x}:J^0(x):     
\nonumber\\\!\!\!\!&=&\!\!\!\!
\sum_\lambda\,\int \! dP\,(\bar{b}(p,\lambda)b(p,\lambda)
-\bar{d}(p,\lambda)d(p,\lambda))\, 
\eea
and
\bea
\label{energy2}
\hat{P}^\mu\!\!\!\!&=&\!\!\!\!
\int\!d^3\bfa{x}:\theta^{0\mu}(x):
\nonumber\\\!\!\!\!&=&\!\!\!\!
\sum_\lambda\,\int \! dP\,p^\mu\,(\bar{b}(p,\lambda)b(p,\lambda)
+\bar{d}(p,\lambda)d(p,\lambda))\,.
\eea

The Feynman propagator, which is defined as the 
time ordered vacuum expectation value
of the field operators,
can be calculated with the results of the preceding equations. 
Adding a small imaginary part to the denominator one finds 
\bea
iS_F(\,x,y\,)_{ij}\!\!\!\!&=&\!\!\!\!\langle\,0\,\vert\,
T\,(\psi_i(x)\,\bar{\psi}_j(y))\,\vert\,0\,\rangle\\
\!\!\!\!&=&\!\!\!\!i\delta_{ij}
\int\! \frac{d^4p}{(2\pi)^4}\,
\frac{e^{-i p_{\mu}( x^\mu-y^\mu)}}{p_{\mu}p^{\mu}-m^2+i\epsilon}
\,,\nonumber
\eea
where the $p^0$ components in the integrand are not
fixed by the mass shell condition, i.e. $p^0\neq \sqrt{\bfa{p}^2+m^2}$. 
From the above expression follows that 
the differential operator of the quantum wave
equation acting on the Feynman propagator is equal to the
four-dimensional delta function 
\be
\label{offsh}
(P\bar{P}-m^2)S_F(x,y)=\delta^4(x-y)\,,
\ee
where the matrix indices are not indicated. 

The consistence
of the above equations is not self-evident if
one uses anticommuting fields. 
The key point, which is responsible for these results, 
is the negative normalization and completeness
of the antiparticle spinors given in Eqs.~(\ref{norma})
and (\ref{compla}). The negative normalization
is also part of the Dirac theory. It is therefore essentially
the same mechanism which provides a consistent treatment
of second quantization in the formalism presented in this paper.

The main intention of this brief discussion was to 
show that a second order differential equation for fermions is in
accordance with basic quantum field theoretical considerations.
In the following, the fields will be considered again
as single-particle wave functions.

\section{Charged massless fermions}
\label{massless}
In the preceding sections
massive electrons have been studied. With
regard to the discussion of photons, 
the $m^2=0$ representation of the Poincar\'e group
will be investigated in the following. 
The discussion is restricted to charged
fermions, i.e.~one can expect that electrons at very high
energy can be described by these states. Furthermore, the free
non-interacting electron field may obey this equation. 
The quantum wave equation for massless fermions has the form 
\be
\label{equat5}
P\bar{P}\psi(x)=0\,.
\ee
Again, the spinor will be defined within the little
group of a standard vector. Since massless particles are moving
with the velocity of light they have no rest frame. Therefore, 
the standard frame will be defined as the system in which
the momentum is directed along the polarization
axis. If the particle is polarized along the z-axis
the positive-energy standard vector is given as 
\be
\label{standml}
p_t^\mu=(\vert p^3\vert,0,0,\, h\vert p^3\vert)\,,
\ee 
where $h=p^3/\vert p^3\vert$ denotes the helicity of the
particle. 

To obtain an arbitrary momentum vector $p^\mu$ one
has to boost perpendicular to the z-axis
\be
\label{rapidy}
p^\mu=(\vert p^3\vert \cosh{\xi},\vert p^3\vert n^1 \sinh{\xi},
\vert p^3\vert n^2\sinh{\xi}, h\vert p^3\vert)\,.
\ee
The unit vector $n^i_\perp=(n^1,n^2,0)$ 
characterizes the direction of the boost. 
With the perpendicular momentum vector $p_\perp^i=(p^1,p^2,0)$
the rapidity is defined by the relation
\be
\label{rapidx}
\tanh{\xi}=\frac{p_\perp}{p^0}\,, 
\ee
where $p_\perp=\vert \bfa{p}_\perp\vert$.
Using $\bfa{\xi}=\bfa{n}_\perp\xi$ the boost can 
be written again in the form $B=\exp{(j\bfa{\xi}/2)}$.
The components of the Pauli-Lubanski
vector in the standard frame of Eq.~(\ref{standml}) are
\bea
\label{PLu}
W^0\!\!\!\!&=&\!\!\!\!-h\vert P_3\vert J^3\nonumber\,,\\
W^1\!\!\!\!&=&\!\!\!\!-\vert P^3\vert(J^1+hK^2)\,,\\
W^2\!\!\!\!&=&\!\!\!\!-\vert P^3\vert(J^2-hK^1)\nonumber\,,\\
W^3\!\!\!\!&=&\!\!\!\!-\vert P^3\vert J^3\nonumber\,.
\eea 
The operators $W^0$ and $W^3$ are linear dependent and will
be represented in the following by $J^3$. 
The three generators $J^3$, $W^1$ and $W^2$ satisfy the
Lie algebra of the Euklidean group in two dimensions 
\be
[\,J^3\,,W^1 ]=iW^2\,,\quad
[\,J^3\,,W^2 ]=-iW^1\,,\quad
[W^1,W^2 ]=0\,,
\ee
which defines the little group of 
the $m^2=0$ representation.
To derive the above equations the angular momentum
operators have to be defined with the orbital
angular momentum operators of Eq.~(\ref{ospinref}).

For the definition of the spin one has to insert 
the generators $J^i=\sigma^i/2$ and $K^i=ij\sigma^i/2$ into Eq.~(\ref{PLu}).
Then one finds in the standard frame and therefore in all
frames
\be
W_{\mu} W^\mu=0\,,
\ee
i.e.~the spin is given in the
degenerate spin $s=0$ representation of $E_2$.
The basis vectors are chosen as eigenvectors 
of $J^3$ with the eigenvalues $\lambda=\pm \frac{1}{2}$.
Therefore, one
can adopt Def.~(\ref{starts1}) with
the Pauli spinor $\vert\lambda\rangle$ and the
standard momentum $p_t^\mu$ of Eq.~(\ref{standml}). A general basis vector
is obtained using Eq.~(\ref{starts2}) with the 
boost parameters defined above.  
This leads to the spinor
\be
\label{m0spnra}
u(\bfa{p},\lambda)=\sqrt{\,\frac{p^0+|p^3|}{2|p^3|}\,}\left(1
+\frac{j\bfa{p}_\perp}
{p^0+\vert p^3\vert}\right)\chi_{\lambda}\,.
\ee
The antiparticle spinor is obtained as in Eq.~(\ref{antis}), i.e.~the
above expression is multiplied
by the hyperbolic unit $j$. 
This ensures that the second quantization for charged
anticommuting fields can be performed consistently 
due to the negative normalization of the spinor
\be
\label{m1spnra}
v(\bfa{p},\lambda)=\sqrt{\,\frac{p^0+|p^3|}{2|p^3|}\,}\left(
\frac{\bfa{p}_\perp}
{p^0+\vert p^3\vert}+j\right)\chi_{\lambda}\,.
\ee
The transformation properties of the basis vectors 
are the same as in Eq.~(\ref{trafog}).

It remains to define the spin operator. A set of four
orthogonal vectors $n^{\mu(\nu)}=(n^{0(\nu)},n^{k(\nu)})$ can be introduced
\be
n^{\mu(0)}=\left(\frac{P^0}{\vert P^3\vert},
\frac{P^k_\perp}{\vert P^3\vert}\right),\quad
n^{\mu(i)}=\left(\frac{P^i_\perp}{\vert P^3\vert},\delta^{ki}+
\frac{P^k_\perp P^i_\perp}{\vert P^3\vert(P^0+\vert P^3\vert)}
\right).
\ee
For the degenerate representation of $E_2$ one operator
is sufficient to characterize the state vectors \cite{Tun85}. 
This operator is chosen as follows
\be
\label{nspin}
S^{3}=\frac{1}{\vert P^3\vert}\, W_\mu\,n^{\mu(3)}\,.
\ee
The spin operator corresponds again to a boosted
generator $J^3$. Explicitly written one finds
\be
S^{3}= \frac{1}{\vert P^3\vert}
(J^3P^0+K^1P^2-K^2P^1)\,.
\ee 

The results can be summarized now.
If one uses the notation
$\vert+\lambda\, P\,\rangle=\vert\, u(\bfa{p},\lambda)\,P\,\rangle$
the particle kets are characterized by the 
operators $\{P_\mu P^\mu, P^\mu, W_\mu W^\mu, S^3\}$
and satisfy the relations
\bea
\label{ndef}
P_\mu P^\mu\,\vert+\lambda\, P\,\rangle
\!\!\!\!&=&\!\!\!\!0\,,
\nonumber\\
P^\mu \,\vert+\lambda\, P\,\rangle
\!\!\!\!&=&\!\!\!\!p^\mu \,\vert+\lambda\, P\,\rangle\,,\nonumber\\
W_\mu W^\mu\,\vert+\lambda\, P\,\rangle
\!\!\!\!&=&\!\!\!\!0\,,
\\
S^3\,\vert+\lambda\, P\,\rangle
\!\!\!\!&=&\!\!\!\!\lambda\,\vert+\lambda\, P\,\rangle
\nonumber\,.
\eea
The plane wave expansion is formally equivalent to
Eq.~(\ref{solute}), but 
the spinors $u(\bfa{p},\lambda)$ and $v(\bfa{p},\lambda)$
have to be replaced 
with the specific $m^2=0$ form
given in Eqs.~(\ref{m0spnra}) and (\ref{m1spnra}). 
The states could also be characterized by the helicity,
whereas in the description presented here the helicity
is fixed and the two polarizations are used to characterize
the two possible states. Changing the chosen helicity  
leads to an interchange of the two polarizations.

\section{The quantum wave equation for photons}
\label{electro}
The motivation for this work was the question whether 
one can find a differential operator which forms the basis
for the description of the fundamental fermion field as well as for the
corresponding gauge field, i.e.~whether there exists an
universal equation of the form
\be
D\phi(x)=m^2_\phi\phi(x)\,,
\ee
where $D$ is a suitable differential operator,
$\phi(x)$ denotes the considered 
quantum field, and $m^2_\phi\geq 0$ corresponds to the mass 
of the particle. 
This means, for fundamental particles the information
about the particle spin should be included only
in the field $\phi(x)$ and not in the differential
operator of the wave equation. 

In quantum electrodynamics, i.e.~more precisely $\phi(x)\in\{\psi(x),A(x)\}$,
this unification is possible, where $D=P\bar{P}$  
has to be chosen as the underlying differential operator. 
It will be shown that the Maxwell equations 
can be derived starting from this unified wave equation. 
For free photon fields one therefore begins with 
\be
\label{wavb}
P\bar{P} A(x)=0\,,
\ee
where $A(x)=A^0(x)+j\bfa{A}(x)$ is a vector field and $m^2_A=0$.
With the electromagnetic fields
\be
\label{emfields}
E^i(x)=-\partial^0A^i(x)-\partial^iA^0(x)\,,\quad 
B^i(x)=\epsilon^{ijk}\partial_jA_k(x)\,,
\ee
one can show that Eq.~(\ref{wavb})
can be expressed
in terms of these fields according to
\bea
\label{maxwell}
P\bar{P} A(x)\!\!\!\!&=&\!\!\!\!-\bfa{\nabla}\cdot\bfa{E}(x)-\partial^0 C(x)\nonumber\\
                  &&\!\!\!\!+ ij\bfa{\nabla}\cdot\bfa{B}(x)\\
                  &&\!\!\!\!-j(\bfa{\nabla}\times\bfa{B}(x)-
                       \partial^0\bfa{E}(x)
                      -\bfa{\nabla}C(x))\nonumber\\
                  &&\!\!\!\!-i(\bfa{\nabla}\times\bfa{E}(x)+\partial^0\bfa{B}(x)
                   )=0\nn\,.
\eea
The calculation has to be done in a specific order. 
In the first step the expression $\bar{P} A(x)$ has
to be evaluated. In the second step the operator
$P$ is acting on the electromagnetic fields.
Calculating $P\bar{P}=P_\mu P^\mu$
is leading to the classical wave equation for the vector potential. 

In the quantum wave equation the four homogeneous 
Maxwell equations are included. 
The straightforward derivation provides 
two additional terms including the field 
\be
\label{zero}
C(x)=\partial_\mu A^\mu(x)\,.
\ee
In the Lorentz gauge this term vanishes and the
correct form of the Maxwell equations is restored.
This indicates that the gauge should not be
chosen arbitrary, i.e.~the vector potential should be perpendicular
to the momentum operator in the relativistic sense 
\be
\label{cond}
P_\mu A^\mu(x)=0\,.
\ee
Otherwise, the six-component tensor structure
of the electromagnetic field would be modified and extended
to a seventh component, which can not be incorporated 
into an antisymmetric second rank tensor. 

In this formalism gauge invariance is not satisfied 
as naturally as in the conventional formulation of 
the Maxwell equations based on
the antisymmetric tensor $F^{\mu\nu}$.
The gauge transformation of the
vector potential $A(x)=A^\mu(x)\sigma_\mu$ is given as usual 
according to
\be
\label{gauge}
A(x)\,\mapsto\,A^\prime(x)=A(x)+\frac{1}{e}\nabla \Lambda(x)\,,
\ee
where $\Lambda(x)$ is a scalar field.
From the electromagnetic fields $\bfa{E}$ and $\bfa{B}$ one knows that they are 
invariant under this transformation. 
Comparing with Eq.~(\ref{maxwell}) one observes that it is only the field $C(x)$
which is required to be invariant. Since this term is supposed
to be zero and the gauge condition should not be modified by
the gauge transformation the scalar field has to fulfil
the relation 
\be
\label{drop}
P\bar{P} \Lambda(x)=0\,.
\ee
Therefore, gauge invariance is achieved in the present formalism
by restricting on a certain class of gauge functions $\Lambda(x)$.

\section{Photon plane wave states}
\label{photop}
In this section a plane wave expansion for free
photon fields will be derived, where the techniques
developed for massless fermions will be applied.
The transformation properties of the vector 
components $A^\mu(x)$ can be understood in terms of 
$4\times 4$ transformation
matrices acting on four-component vectors 
\bea
\label{lorentz2}
&&X^\mu\,\mapsto\, X^{\mu\prime}= (L)^{\mu}_{\,\,\nu}\,X^\nu\,,\nn\\
&&L=\exp{\left(-i(\bfa{J}\cdot\bfa{\theta}+\bfa{K}\cdot\bfa{\xi})\right)}\,.
\eea
For the generators $J^i$ and $K^i$ 
only the third components
are displayed  
\be
(J^3)^{\mu}_{\,\,\nu}=\left(\begin{array}{cccc}
0&0&0&0\\
0&0&-i&0\\
0&i&0&0\\
0&0&0&0\\
\end{array}\right),
\quad
(K^3)^{\mu}_{\,\,\nu}=\left(\begin{array}{cccc}
0&0&0&i\\
0&0&0&0\\
0&0&0&0\\
i&0&0&0\\
\end{array}\right).
\ee
Now, one can use essentially the results of Section \ref{massless}.
The standard frame is again given as the system in which the momentum
is directed along the polarization axis.
The Pauli-Lubanski vector in the standard frame
is given by Eq.~(\ref{PLu}), where the generators now have to
be replaced by the generators of Eq.~(\ref{lorentz2}). 

One can construct two eigenstates $\vert \lambda \rangle$ of 
$J^3$ with the eigenvalues $\lambda=\pm 1$
\be
\vert \pm \rangle=e_\pm^\mu\,\sigma_\mu
=\frac{\mp\, e_1^\mu - ie_2^\mu}{\sqrt{2}}\,\sigma_\mu\,.
\ee
These vectors are eigenstates of $J^3$, 
i.e.~$(J^3)^{\mu}_{\,\,\nu}e_\lambda^\nu=\lambda\,e_\lambda^\mu$. 
$e_1^\mu$ and $e_2^\mu$ are
unit vectors along the x- and y-axis. 
In the standard frame the eigenstates 
satisfy the relation 
\be
W_{\mu} W^\mu \vert \lambda \rangle=(W_{\mu} W^\mu)
^{\alpha}_{\,\,\beta}\,e_\lambda^\beta\,\sigma_\alpha=0.
\ee
The basis vectors of the irreducible
representation are defined according to Eq.~(\ref{starts1}) 
and boosted perpendicular to the polarization axis 
to the system, where the state is described by the momentum $P$ 
\be
\label{starts3}
D(B)\vert\,\lambda\, P_t\,\rangle=
((B)^{\mu}_{\,\,\nu} e_\lambda^\nu)\,\sigma_\mu
\,\vert\,B P_tB^\dagger\,\rangle\,.
\ee
Introducing the abbreviation 
$e^\mu(\bfa{p},\lambda)=(B)^{\mu}_{\,\,\nu}\,e_\lambda^\nu$
the polarization vectors are given as
\be
e^\mu(\bfa{p},\lambda)=\left(\frac{p^\lambda}{\vert p^3\vert},
e_\lambda^i+ \frac{p^i_\perp p^\lambda}{\vert p^3\vert(p^0+\vert p^3\vert)}
\right),
\ee
where $p^\pm=(\mp p^1-
ip^2)/\sqrt{2}$. 
The transformation rules 
for the plane wave states are 
\bea
\label{trafogp}
D(L)\,\vert\,e(\bfa{p},\lambda)\, P\,\rangle\!\!\!\!&=&\!\!\!\!
((L)^{\mu}_{\,\,\nu} e^\nu(\bfa{p},\lambda))\,\sigma_\mu
\,\vert L PL^\dagger\,\rangle\,,
\\
D(T)\,\vert\,e(\bfa{p},\lambda)\, P\,\rangle\!\!\!\!&=&\!\!\!\!
e^\mu(\bfa{p},\lambda)\,\sigma_\mu\,\exp{(-ip_\mu a^\mu)}\vert\, P\,\rangle\,.
\nonumber
\eea
The plane wave states can be written in the 
notation
\be
\vert+\lambda\, P\,\rangle=\vert\, e(\bfa{p},\lambda)\,P\,\rangle
\,,\quad \vert-\lambda\, P\,\rangle=
\vert \,e^{*}(\bfa{p},\lambda)\,-\!\!P\,\rangle \,.
\ee
The spin operator $S^3$
is defined as in Eq.~(\ref{nspin}) with the appropriate
generators for $J^i$ and $K^i$. The defining relations for the
states are formal identical to Eq.~(\ref{ndef}) except  
the different eigenvalues for the polarization. For photons
one finds $\lambda=\pm 1$. 

The plane wave expansion for the
free photon field is given by  
\begin{eqnarray}
\label{solutex}
A(x)\!\!\!\!&=&\!\!\!\!
\sum_\lambda\int \! dP\,\left(\langle\,X^{}
\,\vert+\lambda\, P\,\rangle
\langle\,+\,\lambda\, P\,\vert\,A\,\rangle
+\langle\,X\,\vert-\lambda\, P\,\rangle
\langle\,-\,\lambda\, P\,\vert\,A\,\rangle\right)\\
\!\!\!\!&=&\!\!\!\!\sum_\lambda\int \! \frac{d^3\bfa{p}}{(2\pi)^32p^0}\,\left(
e^\mu (\bfa{p},\lambda)e^{-ip_\mu x^\mu}a(p,\lambda)+
e^{\mu\,*} (\bfa{p},\lambda)
e^{ip_\mu x^\mu}\,\bar{a}(p,\lambda)\right)\sigma_\mu\,.\nn
\end{eqnarray}
The components of the electromagnetic fields can
be obtained from $A(x)=A^\mu(x)\sigma_\mu$.
Since the solution of the quantum wave equation is expanded
in terms of the physical photon states, which are defined 
by the representations of the Poincar\'e group,
the Lorentz gauge condition is satisfied. Therefore,
the $C(x)$ contributions discussed in the last section
vanish and the original form of the
Maxwell equations is restored. 
 
\section{Interactions}
\label{blab1}
As well as in the Dirac theory 
the Lagrange function of the new formalism should be 
invariant under local gauge transformations.
This has the consequence that a gauge field, 
which is identified with the
photon field, has to be introduced by 
a substitution of the momentum operators. The free theory then changes into 
a theory which couples the
electron and photon fields in the interaction terms of the
Lagrangian.  

The gauge field is introduced by replacing 
the momentum operators according to 
\be
P^\mu\,\mapsto\, P^\mu-eA^\mu(x)\,,
\ee
where the charge $e<0$ corresponds to the negative
charge of the electron.
This minimal substitution of the momentum operators 
in the Lagrangian of Eq.~(\ref{Lagr}) is 
leading to 
\bea
\label{Lagrint}
{\cal{L}}(x)\!\!\!\!&=&\!\!\!\!
\bar{\psi}(x)(P-eA(x))(\bar{P}-e\bar{A}(x)
)
\psi(x)\nonumber\\
&&\!\!\!\! -m^2\bar{\psi}(x)
\psi(x)+\frac{1}{2}Tr\left(\bar{A}(x)P\bar{P}
A(x)\right)\,,
\eea
where an additional term for the photon field has been
added to the Lagrangian. 

The Lagrangian (\ref{Lagrint}) is comparable
to the QED Lagrangian for
scalar particles since a seagull term appears which is
not present in the Dirac theory. 
If the gauge field transforms according to Eq.~(\ref{gauge}),
and Eq.~(\ref{drop}) is satisfied,
this Lagrangian is
invariant under a local gauge transformation of
the form
\be
\psi(x)\,\mapsto\,\psi^\prime(x)=e^{-i\Lambda(x)}\psi(x)\,.
\ee
Starting from the above Lagrangian the 
equation of motion for the fermion field can
be calculated using Eq.~(\ref{varaction}) 
\be
\label{basic}
(P-eA(x))(\bar{P}-e\bar{A}(x))\psi(x)=m^2\psi(x)\,.
\ee
For the calculation one should separate 
the basis matrices from the vector components.
The $\sigma_\mu\bar{\sigma}_\nu$-terms,
which are included in the Lagrangian,
are given explicitly in Eqs.~(\ref{spindef}) and (\ref{sigspin}).

In the same way the equation of motion for the photons can
be derived with the Lagrange equation 
\be
\frac{\partial{\cal{L}}}{\partial{A^\mu}}-\partial^{\nu}
\frac{\partial{\cal{L}}}{\partial(\partial^{\nu}A^\mu)}=0\,.
\ee 
This Lagrange equation has been simply 
adopted from the Maxwell theory. 
As well as in the description of the electron
field the spin matrices are considered as
basis vectors which have no further effect on the
dynamical variables. 
Using the above equation a straightforward calculation 
is leading to 
\be
\label{inhom}
P\bar{P}A(x)=-J(x)\,.
\ee
The current $J(x)=J^\mu(x)\sigma_\mu$ includes the photon field and the 
spin matrices $\sigma_\mu\bar{\sigma}_\nu$ between the spinor
functions. Using Eq.~(\ref{spindef}) one can separate the current
into a spin dependent and a spin independent contribution 
\be
\label{crt}
J^\mu(x)=I^\mu(x)+K^\mu(x),
\ee
where the contribution $I^\mu(x)$ is identical to
the electromagnetic current for scalar particles except that two-component
spinor functions are used 
\be
\label{currentx}
I^\mu(x)=
-e\,\bar{\psi}(x)
(\str{P^\mu}\!-\,eA^\mu(x))\psi(x)
+e\,\bar{\psi}(x)(\stl{P^\mu}\!+
\,eA^\mu(x))
\psi(x)\,.
\ee 
The spin dependent part can be expressed with the
spin operators $\sigma^{\mu\nu}$, which are related
to the generators of the
covering group of $SO(3,1)$ according to Eq.~(\ref{spinref2}) 
\be
\label{current}
K^\mu(x)=-
ie\,\bar{\psi}(x)
(\str{P_\nu}\!-\,eA_\nu(x))\sigma^{\nu\mu}\psi(x)
+ie\,\bar{\psi}(x) \sigma^{\mu\nu}(\stl{P_\nu}\!+
\,eA_\nu(x))
\psi(x)\,.
\ee
It is the current given in Eq.~(\ref{crt}) which is conserved when the
electromagnetic field is present. 

Comparing Eq.~(\ref{inhom}) with the Maxwell equations (\ref{maxwell})
one finds, assuming $C(x)=0$, that the inhomogeneous terms are given
in the correct form. 
Eqs.~(\ref{basic}) and (\ref{inhom}) are the new 
equations of motion of quantum electrodynamics.
These equations are very complicated coupled differential equations and
they can be solved only in a suitable approximation scheme. 
To show that these equations provide a reasonable
basis for calculating electromagnetic processes,
the connection of the electron wave equation (\ref{basic}) with the
Dirac equation will be outlined in the following. 

\section{Quantum wave equation and Dirac equation}
\label{blab2}
The Dirac theory
is able to explain experimental data with highest accuracy. 
The quantum wave equation 
should therefore be in agreement with the Dirac equation. 
One can show that in the case of electromagnetical
interactions the differential operator of the
quantum wave equation is equivalent to  
the quadratic form of the Dirac differential operator, 
from which one knows
that the energy spectrum of hydrogen like systems
is exactly the same as for the Dirac equation
\cite{Itz80,Sch93}.

To show this relationship, the quantum wave equation 
(\ref{basic}) will be considered in detail.  
A short calculation is leading to
\be
\label{funda}
\left((P^0-eA^0)^2- (\bfa{P}-e\bfa{A})^2
-j[P^0-eA^0,\bfa{P}-e\bfa{A}]-m^2\right)\psi(x)=0\,,
\ee
where the mass term is now on the left side of the
equation. The second term of the equation
still includes the Pauli
matrices. One can evaluate this expression according to 
\be
(\bfa{P}-e\bfa{A})^2= 
(\bfa{P}-e\bfa{A})\cdot (\bfa{P}-e\bfa{A})-e\bfa{B}\,,
\ee
where one should compare the notation with Eq.~(\ref{produ}).
$\bfa{B}$ corresponds to the magnetic field given in Eq.~(\ref{emfields}).
The commutator, which is proportional to the hyperbolic unit,
can be calculated according to 
\be
[P^0-eA^0,\bfa{P}-e\bfa{A}]=ie\bfa{E}\,,
\ee
with the electric field $\bfa{E}$. Inserting these
results into Eq.~(\ref{funda}) gives 
\be
\label{Pauli}
\left((P-eA)_\mu(P-eA)^\mu -eE^i\,ij\sigma_i+eB^i\sigma_i -m^2\right)\psi(x)=0\,.
\ee
Here, the Pauli matrices $\sigma_i$ are written explicitly for a better
comparision with the Dirac equation.
It is possible to express Eq.~(\ref{Pauli}) completely in
the relativistic tensor formalism if Pauli matrices 
and electromagnetic fields are expressed with the
antisymmetric tensor $\sigma_{\mu\nu}$ given in Eq.~(\ref{sigspin})
and $F^{\mu\nu}=\partial^\mu A^\nu- \partial^\nu A^\mu$
\be
\label{Pauli1}
\left((P-eA)_\mu(P-eA)^\mu -\frac{e}{2}
\sigma_{\mu\nu} F^{\mu\nu} -m^2\right)\psi(x)=0\,.
\ee

This equation is formal identical to 
the quadratic form
of the Dirac equation \cite{Itz80}, which can be derived with
the Dirac formalism. The Dirac equation is given by 
\be
(\gamma_\mu P^\mu- e\gamma_\mu A^\mu(x)-m)\psi(x)=0\,
\ee 
with the Dirac matrices $\gamma_\mu$. 
The quadratic form can be found if one multiplies
the Dirac equation by the operator
$\gamma_\mu P^\mu- e\gamma_\mu A^\mu(x)+m$. This yields
\bea
\label{Pauli2}
(({\gamma}_\mu P^\mu-
 e\gamma_\mu A^\mu)^2-m^2)\psi(x)=&&\nn\\
((P-eA)_\mu(P-eA)^\mu-\frac{i}{2}\sigma_{\mu\nu}
[P^\mu-eA^\mu,P^\nu-eA^\nu]-m^2)\psi(x)=&&\nn\\
((P-eA)_\mu(P-eA)^\mu -\frac{e}{2}
\sigma_{\mu\nu} F^{\mu\nu} -m^2)\psi(x)=&\!\!\!\!0\,.&
\eea

The two wave equations of
Eqs.~(\ref{Pauli1}) and (\ref{Pauli2}) have the same
form. However, there are two differences:
The first difference 
is given in the structure of the spinors $\psi(x)$. In the
case of the quantum wave equation $\psi(x)$ has a two-component structure, 
whereas in the Dirac equation $\psi(x)$ corresponds to a four-component spinor 
\bea
\mathrm{Quantum\,wave\,equation:}\hspace{0.5cm}
\psi(x)\!\!\!\!&=&\!\!\!\!\varphi(x)+j\chi(x)\,,\nn\\[0.5\baselineskip]
\mathrm{Dirac\,equation:}\hspace{0.5cm}
\psi(x)\!\!\!\!&=&\!\!\!\!\left(\begin{array}{c}
\varphi(x)\\ \chi(x)
\end{array}\right).
\eea
The other difference is the spin tensor $\sigma_{\mu\nu}$. 
In the Dirac theory this term is defined 
according to $\sigma_{\mu\nu}=i/2\,[\gamma_\mu,\gamma_\nu]$.
With this tensor one is able to express Eq.~(\ref{Pauli2}) 
according to
\be
\label{Pauli3}
\left((P-eA)_\mu(P-eA)^\mu -eE^ii\alpha_i+eB^i\sigma_i -m^2\right)\psi(x)=0\,.
\ee

Comparing this equation with Eq.~(\ref{Pauli}) one observes
that in both cases the term including the electric field is the only
term which couples the upper and the lower component
of the spinor. In the quantum wave 
equation the coupling term
is proportional to $j\sigma_i$, in the quadratic Dirac equation
the term corresponds to $\alpha_i=\gamma_0\gamma_i=\gamma_5\sigma_i$.
One can show, using the Dirac
representation of $\gamma_5$, that $j$ and $\gamma_5$ have
the same effect on the spinor, an interchange between upper and
lower components
\bea
\mathrm{Quantum\,wave\,equation:}\hspace{0.48cm}
\it j\psi(x)\!\!\!\!&=&\!\!\!\!\chi(x)+j\varphi(x)\,,\\[0.5\baselineskip]
\mathrm{Dirac\,equation:}\hspace{0.3cm}
\gamma_5\psi(x)\!\!\!\!&=&\!\!\!\!
\left(\begin{array}{cc}
0&1\\
1&0\\
\end{array}\right)\left(\begin{array}{c}
\varphi(x)\\ \chi(x)
\end{array}\right)=\left(\begin{array}{c}
\chi(x)\\ \varphi(x)
\end{array}\right).\nn
\eea
One therefore finds in both cases the same two coupled
differential equations. In the quantum wave equation
the terms proportional to the hyperbolic unit
belong to one differential equation, the other terms to
the second equation. In the quadratic Dirac equation
the differential equations are separated by
the component structure.

\section{Summary and Conclusions}
In this work a new formulation of quantum electrodynamics 
was presented. The motivation for the investigation was
the assumption that fundamental free quantum fields
should be described by a unified differential operator, which does
not depend on any particle properties.
In the case of electrons and photons it was 
shown that, concerning the spin
structure, this unification is possible. The wave equation constructed
with this differential operator
has been denoted by \textit{quantum wave equation}. 
It is leading to the same
coupled differential equations for electrons 
as the quadratic form of the Dirac equation. 
Furthermore, the quantum wave equation for photons is equivalent
to the four Maxwell equations.

These results can be obtained if the basis vectors of the
relativistic vector coordinates are represented by
a relativistic matrix algebra which includes
the unit matrix and the Pauli matrices multiplicated by the 
hyperbolic unit. 
Relativistic vectors based on this matrix algebra
were investigated. Their behaviour
under Lorentz transformations
was studied and a relativistic spin group was constructed
in analogy to the non-relativistic treatment.
The quantum wave equation for photons shows that 
vector fields appear in combination with the
same matrix algebra. Therefore, it seems to be
a general property of quantum physics that
vector coordinates have to be considered together 
with the corresponding algebra for the basis vectors.

It was shown that the
differential operator of the quantum wave equation 
transforms like a spin operator.
However, for a non-interacting system the spin structure of the
differential operator is not important
and the operator is reduced to the mass operator of the Poincar\'e group. 
Therefore, the properties of the free electron field
have been investigated within this group and  
the plane wave representation was generated explicitly.
It was shown that a boosted Pauli matrix 
can be related to the Pauli-Lubanski vector.
This offers the possibility to define
the spin in analogy to the non-relativistic treatment.
Former problems in the second quantization of anticommuting 
Klein-Gordon fields disappear if the
negative energy contribution is multiplicated 
by the hyperbolic unit.
A free massless charged fermion field has been investigated, where 
the plane wave expansion of the field can be constructed in analogy to the
massive electron field. 

The quantum wave equation for photons
is leading to the Maxwell equations including 
two additional terms. These terms vanish in the Lorentz gauge. 
A plane wave expansion for the photon field was derived,
which removes the additional terms and stays in close
analogy to the description of the electrons. 

Though the spin structure is not important in a non-interacting system, 
the inclusion of interactions requires 
the given form of the quantum wave equation. A new Lagrange function
of quantum electrodynamics
can be constructed using the concept
of minimal substitution. The equations of
motion for the electron and photon fields were derived
and the relation of the electron equation
to the quadratic form of the Dirac equation was shown. 
The equation of motion for the photons is equivalent to the
inhomogeneous Maxwell equations. 
Since the new formalism can be related with the Dirac
and the Maxwell equations one can expect
that explicit calculations of observables will lead
to results which should be close to the results of
the conventional formalism.


\begin{thebibliography}{99}
\bibitem{Dir28}
P. A. M. Dirac, Proc. Roy. Soc. (London) \textbf{A 117} (1928),
610.
\bibitem{Hes66}
D. Hestenes, "Space Time Algebra",
Gordon and Breach, New York, 1966.
\bibitem{Ban95}
W. E. Baylis, ed., "Clifford (Geometric) Algebras
with applications to physics, mathematics, and engineering",
Birkh\"auser, Boston, 1996.
\bibitem{Dur35}
A. Dura\~nona Vedia and J. C. Vignaux,
Publ. de Facultad Ciencias
Fisiciomatematicas Contrib. (Universidad Nacional de La Plata -
Argentina) \textbf{104} (1935), 139.
\bibitem{Cap41} P. Capelli, 
Bull. of American Mathematical Society {\bf 47} (1941), 585.
\bibitem{Sor79}
L. Sorgsepp and L. Lohmus,  Hadronic J. \textbf{2} (1979),
1388.
\bibitem{Fje86}
P. Fjelstad,  Am. J. Phys. \textbf{54} (1986),
416.
\bibitem{Hes91}
D. Hestenes, P. Reany and G. Sobczyk,
Ad. Appl. Cliff. Alg. \textbf{1} (1991),
51. 
\bibitem{Kel94}
J. Keller, Ad. Appl. Cliff. Alg. \textbf{4} (1994), 1.
\bibitem{Ant98}
F. Antonuccio, hep-th/9812036 (1998).
\bibitem{Tun85}
Wu-Ki Tung, "Group Theory in Physics",
World Scientific, Singapore, 1985.
\bibitem{Wig39}
E. P. Wigner, Annals of Mathematics \textbf{40} (1939),
149.
\bibitem{Bjo65} J.D. Bjorken and S.D. Drell,
                "Relativistic Quantum Fields",
                McGraw-Hill, New York, 1965.
\bibitem{Ryd85}
L. H. Ryder, 
"Quantum Field Theory",
Cambridge University Press, Cambridge, UK, 1985.
\bibitem{Mic59}
L. Michel, Il Nouvo Cimento Supplemento \textbf{14} (1959), 95.
\bibitem{Wig60}
A. S. Wightman, in 
"Relations de Dispersions et Particules \'El\'ementaires",
(C. DeWitt and M. Jacob, Eds.),
Hermann and John Wiley, New York, 1960. 
\bibitem{Itz80} C.~Itzykson and J.-B.~Zuber, "Quantum
                 Field Theory", McGraw--Hill, New York, 1980.
\bibitem{Sch93}
G. Scharf, "Finite Quantum Electrodynamics", Springer, Berlin, 1993.
\end{thebibliography}
\end{document}